\documentclass[12pt]{article}
\usepackage{amsmath,amssymb}
\usepackage{eucal}

\usepackage[dviwindo]{graphicx}
\newcommand{\bs}{\begin{subequations}}
\newcommand{\es}{\end{subequations}}
\numberwithin{equation}{section}
\def \myfigures #1#2#3#4
{\begin{figure}[ht]
    \begin{center}
        \includegraphics[width=#2 \textwidth]{#1.eps}
        \hfill
        \includegraphics[width=#4 \textwidth]{#3.eps}
    \end{center}
\end{figure}}

\newcommand{\ben}{\begin{eqnarray}}
\newcommand{\een}{\end{eqnarray}}
\newcommand{\la}{\label}
\begin{document}
%\DeclareGraphicsExtensions{.jpg,.pdf,.mps,.png}

\title{Novel Properties of Bound States of Klein-Gordon Equation in Gravitational Field of Massive Point}

\vskip 1.5truecm

\author{P.P. Fiziev\thanks{Department of Theoretical Physics, University of Sofia,
E-mail:\,\,\,fiziev@phys.uni-sofia.bg},
T.L. Bojadjiev\thanks{LCT, JINR, Dubna, Russia,
E-mail:\,\,todorlb@fmi.uni-sofia.bg}, D.A. Georgieva
\thanks{Department of Theoretical Physics, University of
Sofia, E-mail: dgeorgieva@phys.uni-sofia.bg}}

\date{}
\maketitle

\begin{abstract}
We consider for the first time the solutions of Klein-Gordon
equation in gravitational field of {\em massive} point source in
GR. We examine numerically the basic bounded quantum state and the
next few states in the discrete spectrum for different values of
the orbital momentum. A novel feature of the solutions under
consideration is the essential dependence if their physical
properties on the gravitational mass defect of the point source.
Such mass defect was not introduced up to recently. Its variation
yields a repulsion or an attraction of the quantum levels up to
their quasi-crossing.
\end{abstract}

%%%%%%%%%%%%%%%%%%%%%%%%%%%%%%%%%%%%%%%%%%%%%%%%%%%%%%%%%%%%%%%%%%%
%\draft
\sloppy
%\scrollmode
%%%%%%%%%%%%%%%%%%%%%%%%%%%%%%%%%%%%%%%%%%%%%%%%%%%%%%%%%%%%%%%%%%%%

\section{Introduction}
The correct solutions of the Einstein equations for the
gravitational field of point particle with bare mechanical mass
$M_0$ were found recently \cite{Fiziev}. It turns out that these
{\em fundamental}  solutions are described by mathematical
distributions and own the necessary jumps in the first derivatives
of the metric, needed to satisfy the Einstein equations at the
place of the point source. The new solutions form a two-parameters
family of metrics on singular manifolds
$\mathbb{M}^{(1,3)}\{g_{\mu\nu}\}$. They are defined by the bare
mass $M_0$ and by the Keplerian mass  $M$, or equivalently, by the
Keplerian mass $M$ and the ratio of the masses:
$\varrho=M/M_0\in(0,1)$. This ratio describes the gravitational
mass defect of the point particle.

The mathematical and the physical properties of the new solutions
are essentially different in comparison with the well known other
spherically symmetric static solutions to the Einstein equations
with different type of singularities at the center of the
symmetry, which is surrounded by an empty space. The previously
known solutions were often considered as a solutions for
describing of single point particle in GR. In the most of the
known cases this turns to be incorrect \cite{Fiziev}. Therefore
the new solutions call for reconsideration of many of commonly
accepted notions and their physical interpretation.

Here we are starting the study of the GR wave equations in the
gravitational field of massive point. In the present article we
shall consider as an example only the simplest GR wave equation --
the Klein-Gordon one. We shall present a numerical study of this
well known equation which describes (anti)particles of zero spin
\cite{FV}.

The reader can find the basic results of the analysis of this
equation in the {\em vacuum} Schwarzschild metric, when described
in the commonly used Hilbert gauge or in other gauges, for
example, in the recent review article \cite{Jacobson}, in
\cite{QNM}, and in the large amount of the references therein. A
discussion of different radial gauges and corresponding references
one can find in \cite{Fiziev}.

\section{The Formulation of the Problem}

Making use of the natural radial variable  of the problem:
$r\in(0,\infty)$ \cite{Fiziev}, one can represent the new {\em
regular} solutions for the gravitational field outside of the
massive point source in the form:
\ben \la{New_metric}
    ds^2 = e^{2\varphi_{\!{}_G}} \left[dt^2 - {{dr^2}\over{N_{\!{}_G}(r)^4}}\right] -
    \rho(r)^2 \left(d\theta^2\!+\!\sin^2\!\theta d\phi^2\right)\!.\hskip .2truecm
\een
Here
\ben
    \varphi_{\!{}_G}(r;M,M_0):=-{{G_N M}\over {r+G_N M}/\ln({M_0\over M})} \la{Gpot}
\een
is the modified Newton potential, the coefficient
$N_{\!{}_G}(r)=\left(2\varphi_{\!{}_G}\right)^{-1}
\left(e^{2\varphi_{\!{}_G}}-1\right)\,,$
the Hilbert luminosity variable is
\ben \rho(r) = \frac{2G_N M}{1 - e^{2\varphi_{\!{}_G}}}=
{{r\!+\!G_NM/\ln({M_0\over M})}\over {N_{\!{}_G}(r)}}\,, \la{rhoG}
\een
and $G_{{}_N}$ is the Newton gravitational constant.

In Hilbert gauge, outside of the source, the solution has a
standard form: \ben
g_{tt}(\rho)\!=\!1\!-\!{{\rho_G}/{\rho}},\,\,\, g_{\rho\rho}(\rho)
= -1/g_{tt}(\rho), \la{HilbertSol} \een where $\rho_G =
2G_{{}_N}M$ is the Schwarzschild radius. We have to stress that
the presence of the matter source forces us to consider this form
of the solution {\em only } on the physical interval of the
luminosity variable $\rho\in (\rho_0,\infty)$, where
\ben
    \rho_0 = 2G_N M /(1-\varrho^2)\geq \rho_G.\la{rho_0}
\een

One can find in \cite{Fiziev} the physical and the mathematical
justification of this {\em cutting procedure} of the admissible
interval of luminosity variable  $\rho$ for the discovered by
Hilbert, Droste and Weyl \cite{HDW} form of the solution
\eqref{HilbertSol}. Here we shall stress that such unusual cutting
is a ultimate consequence of the existence of a {\em matter}
source of the gravitational field. After all, it permit us to
satisfy the physical requirement not to allow an infinite
luminosity of the point sources in correspondence to the real
observations. In the articles \cite{Fiziev} it was shown that in
GR we have a natural cutting parameter for the classical
divergences. It originates from the drastic change of the
space-time geometry due to the infinite mass-energy density of the
point particle. It turns out that in presence of matter source all
phenomena, related with the event horizon, together with the very
horizon, belong to the {\em non-physical} domain of the variables.
Therefore, in accord to Dirac's intuition \cite{Dirac}, all such
notions "should not be taken into account in any physical theory"
of matter.

The luminosity variable $\rho$ is an exclusive one, because the
results, expressed by making use of it, are {\em locally}
invariant under radial gauge transformations. From computational
and from physical point of view a more preferable variable is: $g
= g_{tt} = 1 - \rho_G/\rho$. Then \ben ds^2 = g\,dt^2\!-\!\rho_G^2
\left[{{dg^2}\over{g(1-g)^4}}+{{d\theta^2\!+\!\sin^2\!\theta\,
d\phi^2} \over{(1-g)^2}}\right]. \la{dsg} \een

In these variables the 4D D'Alembert operator in the curved
space-time $\mathbb{M}^{(1,3)}\{g_{\mu\nu}\}$, created by a
massive point, acquires the form
\ben \la{Dalamber}
    \square = g^{-1}\partial_t^2-\rho_G^{-2}\Big((1-g)^4\partial_g\left(g\partial_g\right)+
(1-g)^2\Delta_{\theta\phi}\Big) \een where
$\Delta_{\theta\phi}=\sin^{-1}\!\theta\,\partial_\theta\left(\sin\theta\,\partial_\theta\right)
+\sin^{-2}\!\theta\,\partial_\phi^2$.

The Klein-Gordon equation for test particles of mass $m$ in
space-time $\mathbb{M}^{(1,3)}\{g_{\mu\nu}\}$ has a standard form:
\ben \la{KG}
    \square \Phi + m^2\Phi=0.
\een

As a result of the spherical symmetry the pseudo-Reimannian
manifold $\mathbb{M}^{(1,3)}\{g_{\mu\nu}\}$, created by the point
particle, has a group of motion $SO(3)$. Nontrivial are the
quantities and equations  only on the quotient space
$\mathbb{M}^{(1,1)}=\mathbb{M}^{(1,3)}/SO(3)$, with natural global
coordinates $t$ and $g$. The reduction of Eq. \eqref{KG} on the
quotient space $\mathbb{M}^{(1,1)}$ can be reached by the
substitution
$\Phi(t,g,\theta,\phi)=\Psi_l(t,g)Y_{l,l_z}(\theta,\phi)$, where
$Y_{l,l_z}(\theta,\phi)$ are the standard spherical functions:
$\Delta_{\theta\phi}Y_{l,l_z}(\theta,\phi)=-l(l+1)Y_{l,l_z}(\theta,\phi)$,
à $l=0,1,2,...$; $l_z=-l,...,0,...,l$.

For the time-dependent radial wave function $\Psi_l(t,g)$ on the
manifold $\mathbb{M}^{(1,1)}$ we have the partial differential
equation:
\ben g^{-1}\partial_t^2\Psi_l -\rho_G^{-2} \left\{(1-g)^4
\partial_g \left( g \partial_g \Psi_l \right) +
\left[\,l(l+1)(1-g)^2 + m^2 \right] \Psi_l\right\} = 0. \la{Psi}
\een

The space-time $\mathbb{M}^{(1,3)}\{g_{\mu\nu}\}$ of the problem
at hand is invariant under translations with respect to the time
variable $t$, as well. This symmetry of
$\mathbb{M}^{(1,3)}\{g_{\mu\nu}\}$ is related to the translation
group $T(1)$: $t\to t+ const$ and allows the total reduction of
the problem to one dimensional one using the anzatz
$\Psi_l(t,g)=e^{-iEt}R_l(g)$. For the time-independent radial wave
function $R_l(g)$ on the quotient space
$\mathbb{M}^{(1)}=\mathbb{M}^{(1,1)}/T(1)$ we obtain an ordinary
differential equation:
\ben\la{R} {{d^2R_l}\over{dg^2}}+{1\over g} {{dR_l}\over{dg}} +
\left[{{\varepsilon^2} \over{g^2(1-g)^4}}-
{{\mu^2}\over{g(1-g)^4}}- {{l(l+1)}\over{g(1-g)^2}}\right] R_l=0.
\een
Here $\varepsilon = \rho_G\,E$ and $\mu=\varrho_G\,m$ are,
correspondingly, the dimensionless total energy and mass of the
spinless test particles. We are using the units $c=\hbar=1$. The
dimensionless orbital momentum in our units is $l=L/m\rho_G$,
where $L$ is the dimensionfull one.

\begin{figure}[ht] \label{F0}
\begin{center}
        \includegraphics[width=0.48 \textwidth]{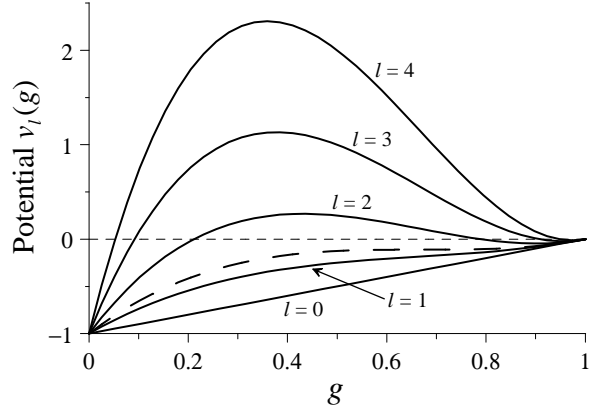}
        \hfill
        {\caption
        {The potential $v_l(g)$ for
        $l=0, 1, 2, 3, 4$ and $\mu=1$. The dashed line corresponds to
        the value $l^{crit}_v={1\over 2}\sqrt{1+12 \mu^2}-{1\over 2}$ for which
        the maximum of the function $v_l(g)$ arise.}}
%        \hfill
%        \parbox[t]{#8\textwidth}{\caption {#7}\label{#5}}
    \end{center}
\end{figure}

The motion of the Klein-Gordon particles with a fixed orbital
momentum $l$ can be treated as a relativistic motion in a
dimensionless potential
\ben v_l(g)=(1-g)\left[l(l+1)g(1-g)-\mu^2\right], \,\,\,g\in
[g_0,1],\la{vg} \een
which is shown (for $\mu=1$ and several different values $l\geq
0$) in Fig.~\ref{F0}. The point $g=0$ corresponds to the event
horizon and is placed in the nonphysical domain $g\in [0, g_0)$,
where $g_0=\varrho^2>0$. The point $g=1$ corresponds to the
physical infinity (with respect to the variables $r$, or $\rho$).
There the space-time $\mathbb{M}^{(1,3)}\{g_{\mu\nu}\}$ is flat.
Writing down the Eq. (\ref{R}) in the form
$$(1-g)^4\left(g\frac{d}{dg}\right)^2R_l + \left[\,\lambda\!-\!v_l(g)\right]\, R_l = 0$$
we see that the points $g=0$  è $g=1$ are singular points of this
equation. The second one is a non-regular singular point. Here
$\lambda=\varepsilon^2-\mu^2$.

The substitutions:
\begin{equation} \la{uP}
    u=\ln \left({g\over {1-g}}\right) + {1\over (1-g)},
\end{equation}
$$R_l(g)=(1-g)P_l(g)\,,$$
introduce ``the tortoise coordinate'' $u$ and the corresponding
radial wave function  $P_l(u)$, and lead to the following standard
(Schr\"odinger like) form of the Eq.~(\ref{R}):
\ben \la{Pl}
    \frac{d^2 \,P_l}{du^2} + \left[\,\lambda - w_l(u)\right]\!P_l=0\,.
\een\
The bounded quantum states can be found as a solutions of
\eqref{Pl} under the boundary conditions
\ben\la{boundary}
    P_l(u_0)=0,\quad P_l(\infty) = 0\,,
\een
together with the usual $L_2 $-normalization
\ben \la{norm}
    \int\limits_{u_0}^\infty P_l^2(u)\,du - 1 = 0\,.
\een  The quantity
\ben \la{wl}
    w_l(g)= v_l(g) + g(1-g)^3=(1-g)\left[l(l+1)g(1-g)+g(1-g)^2-\mu^2\right]
\een
is the extended dimensionless potential of the problem and
depends implicitly on the variable $u$ via the solution \eqref{uP}
of the Cauchy problem
\ben \la{geq}
    \frac{d\,g}{du} = g\,(1-g)^2\,, \quad g(u_0)=\varrho^2\,.
\een

An essential novel feature of our approach to the problem is the
correct fixing of the physical domain of the variables: the
potentials $v_l$ and $w_l$ must be considered only on the interval
$[u_0, \infty)$, where $u_0 = u_0(\varrho^2)$. Thus the
nonphysical infinitely deep potential well at $\rho\to 0$, and the
event horizon at $\rho=\rho_G \left(\Rightarrow
\!u_G\!=\!-\infty\right)$, are excluded from our consideration.
\begin{figure}[here]
    \begin{center}
        \includegraphics[totalheight=6.5cm,keepaspectratio]{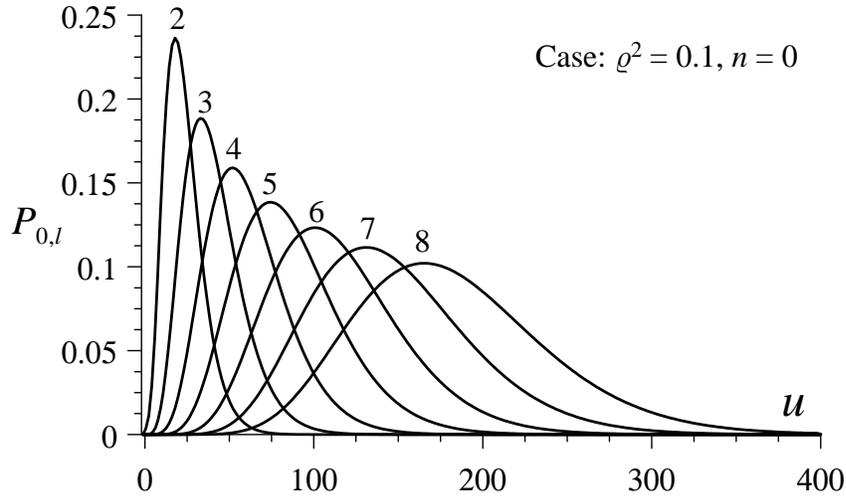}
        \caption{The radial wave function $P_{0\,l}(u,\varrho^2)$ for different
        values of the quantum number $l=2, 3, 4, 5, 6, 7, 8$.} \label{F1}
    \end{center}
\end{figure}

As a completely new phenomenon, a discrete spectrum comes into
being in the problem at hand. We denote by
$\varepsilon_{n\,l}=\sqrt{\lambda_{n\,l}+\mu^2}$ the discrete
values of the energy $\varepsilon$ of the bounded states. The
principle quantum number $n=0,1,...$ determines the number of the
zeros of the radial wave function $P_{n\,l}(u)$ in the interval
$u\in (u_0,\infty)$ and of the function
$R_{n\,l}(g)=(1-g)P_{n\,l}\big(u(g)\big)$ -- in the interval $g\in
(g_0,1)$.

\section{Basic Numerical Results}

\begin{figure}[ht]\vspace{.5truecm}
    \begin{center}
        \includegraphics[totalheight=6.0cm,keepaspectratio]{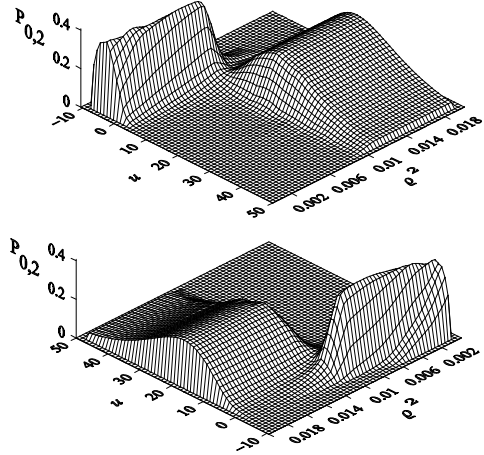}
        \caption{The radial wave function $P_{0\,2}(u;\varrho^2)$ for
        different values of the two variables $u$ and the squared mass ratio
        $\varrho^2$.} \label{F2}
    \end{center}
\end{figure}

\begin{figure}[ht]
    \begin{center}
        \includegraphics[totalheight=5.7cm,keepaspectratio]{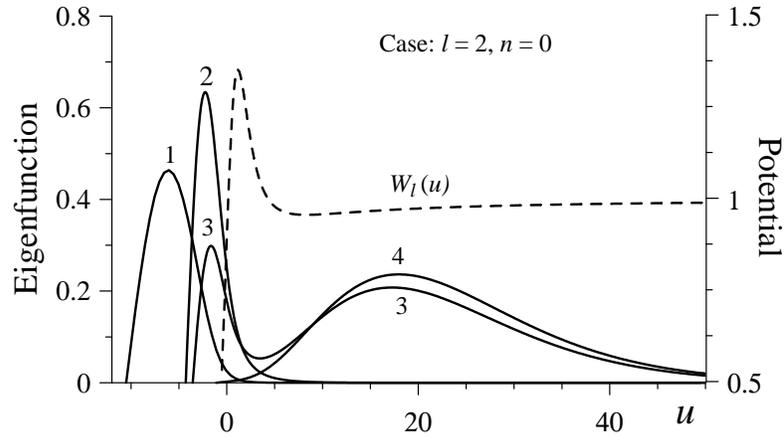}
        \caption{The radial functions $P_{0\,2}(u;\varrho^2)$ for different
        values of the gravitational mass defect of the source.} \label{F3}
    \end{center}
\end{figure}

For the numerical study of the solutions of the Sturm-Liouville
problem \eqref{Pl} -- \eqref{norm}  on some finite interval
$\left[ u_0, u_\infty \right]$ we use the continuous analog of
Newton method \cite{pazppsl_99}. The corresponding linear boundary
problems at each iteration are solved by means of
spline-collocation scheme of fourth order of approximation. The
convergence of the calculated eigenfunctions and corresponding
eigenvalues $\lambda_{n\,l}$ for large enough values of $u_\infty$
is proved on a set of embedded intervals. As a result of numerical
experiments we have obtained a number of new properties of the
bounded quantum states of the problem at hand, a part of which
were unexpected to some extend.

The basic results for the radial wave function $P_{n\,l}(u)$ for
different values of the orbital quantum number $l$ an $n=0$ are
shown in Fig .~\ref{F1}.

In the Fig.\ \ref{F2}, using as an example the wave function
$P_{0\,2}(u;\varrho^2)$, we show the typical dependence of the
wave functions of the bounded states both on the variables $u$ and
on the squared mass ratio $\varrho^2$.

The radial functions $P_{0\,2}(u;\varrho^2)$ for different values
of the gravitational mass defect $\varrho$ are shown in
Fig.~\ref{F3}. As seen, depending on the values of the squared
mass ratio $\varrho^2$ of the point source, the wave function of
the test particle in the point particle's gravitational field is
located in the inner potential well (for small $\varrho^2$ -- the
curves 1 and 2 in Fig.~\ref{F3}), or in the exterior potential
well (for big $\varrho^2$ -- the curve 4 in Fig.~\ref{F3}). In the
narrow intermediate domain of values of $\varrho^2$ a transition
regime take place -- the curve 3 in Fig.~\ref{F3}. Actually, this
transition depends continuously on the variable $\varrho^2$, but
it seems that it is a jump-like, because the transition happens in
a quite smaller scales with comparison to the other much more
smooth changes of the corresponding quantities. (See, for example,
the dependence of the eigenvalues on the variable $\varrho^2$).
These phenomena are described in a more transparent way in the 3D
Fig.~\ref{F2}.

In our approach to the problem the basic new physical phenomena
are related with the gravitational mass defect of the point
source. Such mass defect was not considered and studied until now,
because for the standard Hilbert form of the Schwarzschild
solution "the bare rest-mass density is never even introduced"
\cite{ADM} correctly.

The dependence of the first four eigenvalues $\lambda_{0\,l, 1\,l,
2\,l, 3\,l}$  on the squared mass ratio $\varrho^2$  is shown in
Fig.~\ref{F5} for $l=2$. A typical steep-like dependence of the
discrete eigenvalues $\lambda_{n\,l}(\varrho^2)$ on the variable
$\varrho^2$ is seen. For each value of the principle quantum
number $n$ the number of the jumps of the energy levels
$\varepsilon_{n\,l}(\varrho^2)=\sqrt{\lambda_{n\,l}(\varrho^2)+\mu^2}$
depends on the number of the maxima of the corresponding wave
function which are moved from the inner well to the exterior one
during the transition regime which develops with the increase of
the values of the mass ratio $\varrho^2$.

\vskip .5truecm
\myfigures{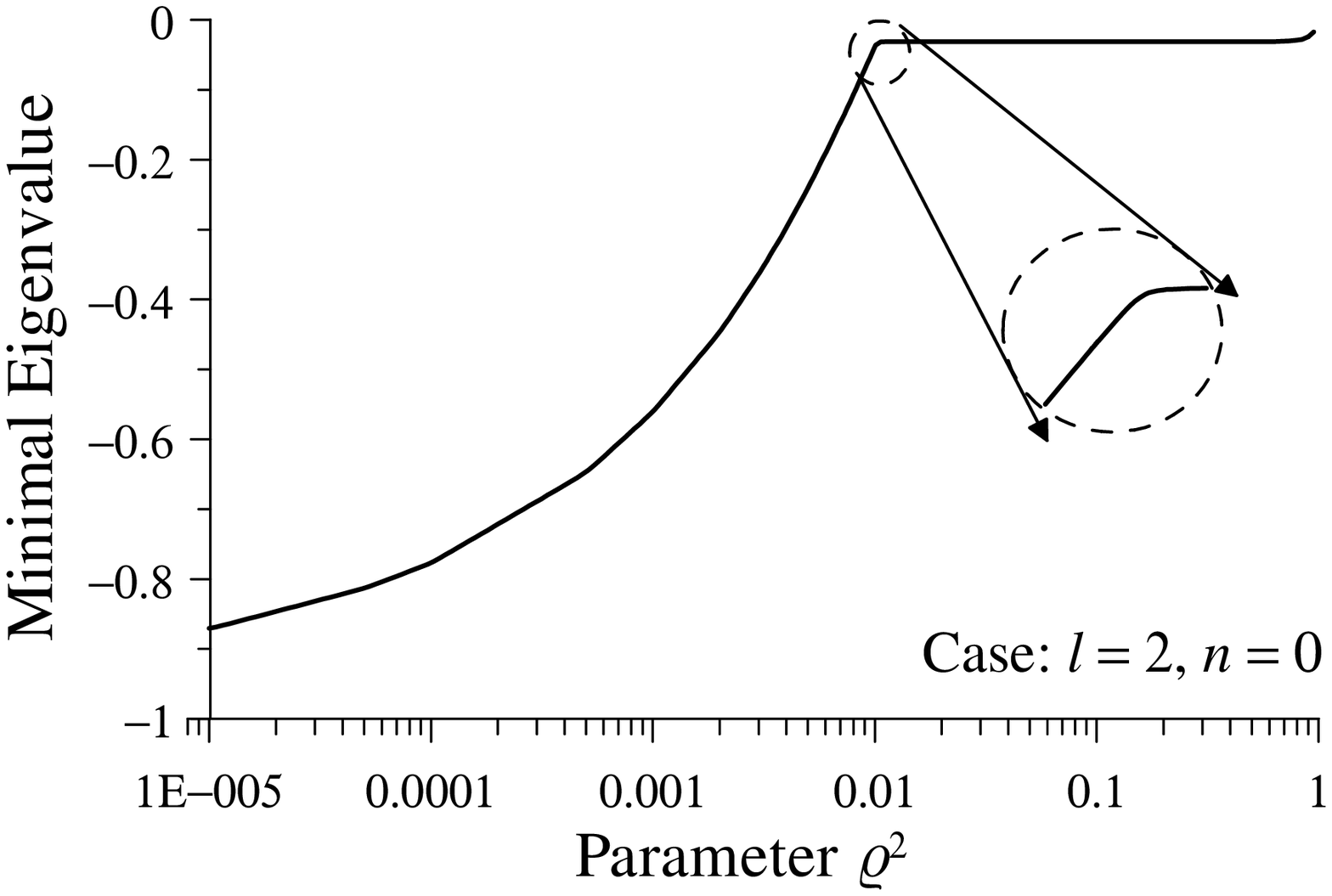}{0.48}{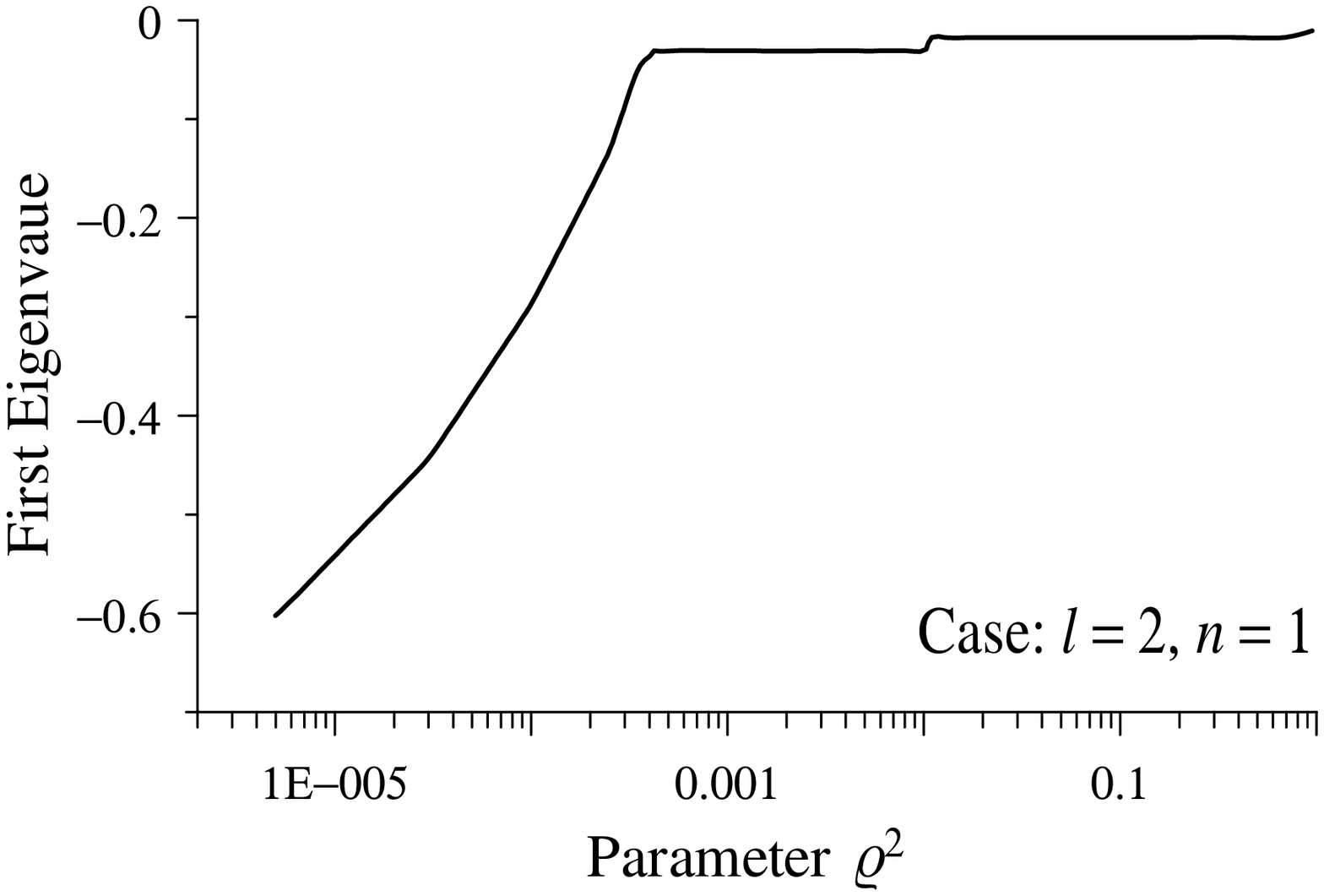}{0.48} \myfigures{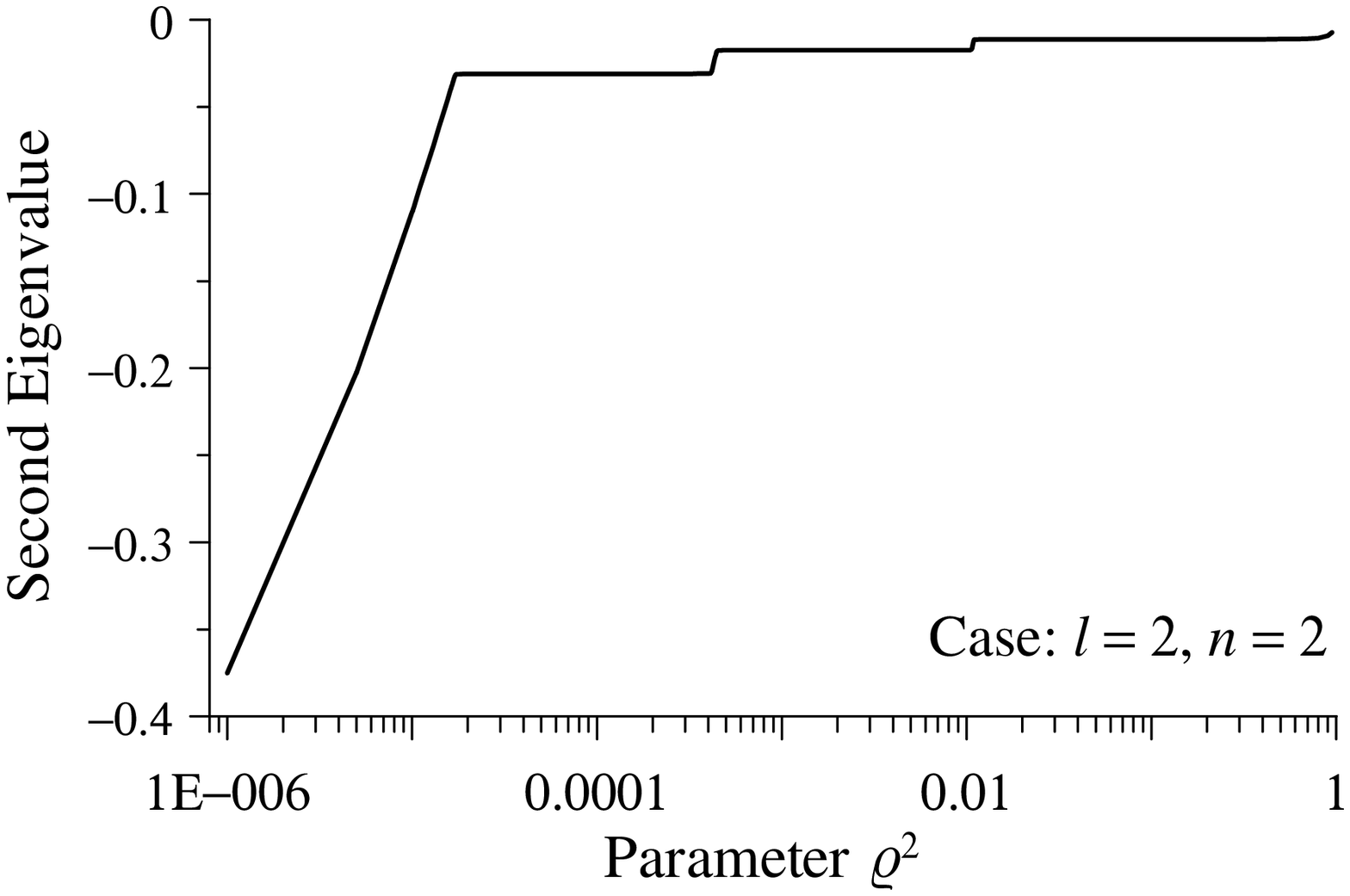}{0.48}{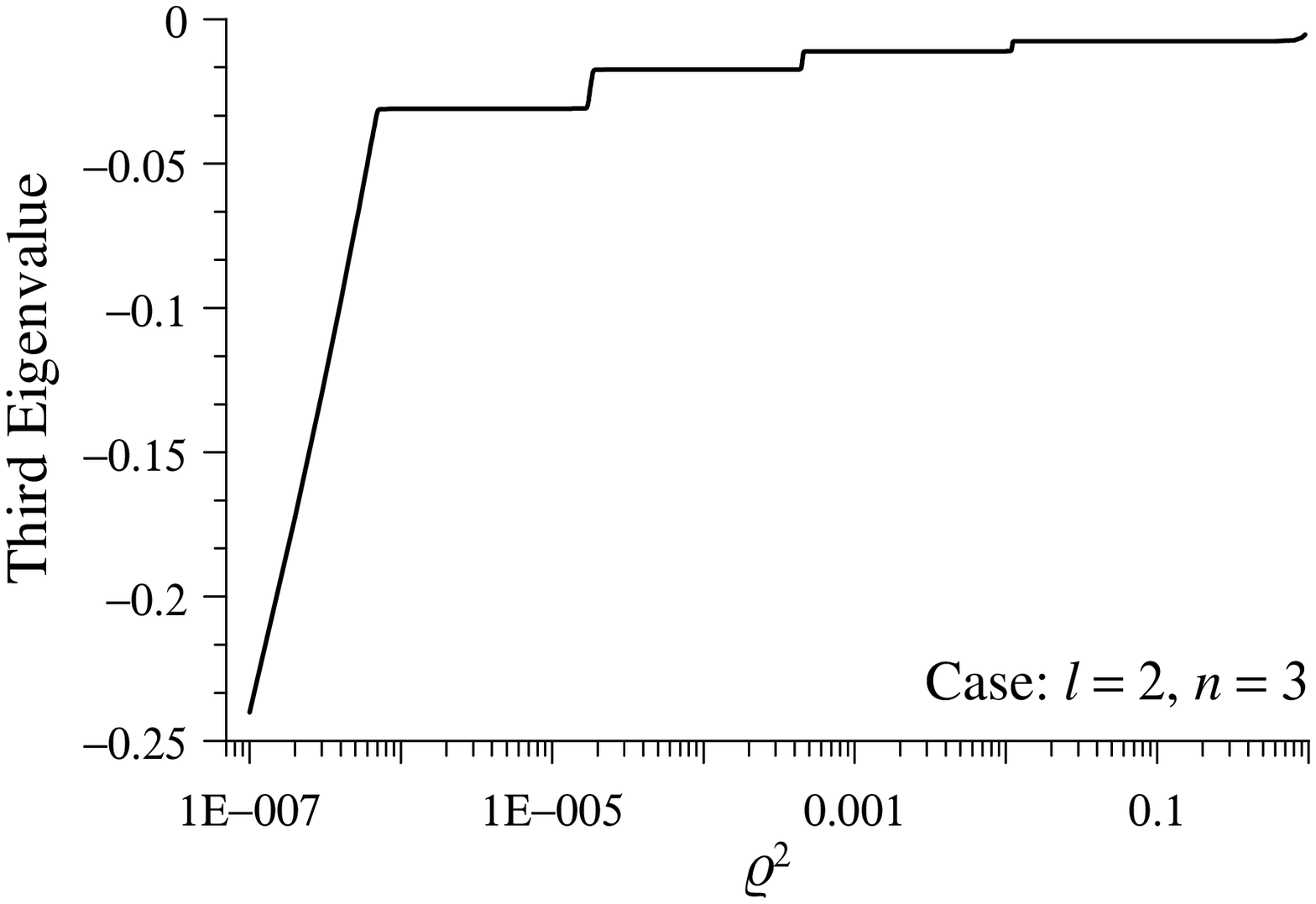}{0.48}
\begin{figure}[ht]
        \caption{The dependence of the first four eigenvalues
$\lambda_{0\,l, 1\,l, 2\,l, 3\,l}$  on the mass ratio
$\varrho^2.$} \label{F5}
\end{figure}

It's clear that such nontrivial behavior of the energy levels
$\varepsilon_{n\,l}(\varrho^2)$ of the Klein-Gordon equation in a
gravitational field of a point particle is due to the existence of
two {\em finite} potential wells: 1) An inner one which has a size
of order of the Schwarzschild radius and in general case of
macroscopic orbital momenta is very deep; and 2) An exterior one,
which is extremely  shallow in comparison with the inner well. The
normal world with almost Newton gravity "lives" in the exterior
well. The both wells are separated typically by a huge potential
barrier which looks like a centrifugal barrier from outside, and
as a suspensory barrier of the type of a spherical potential wall
from inside. Our calculations describe the quantum penetration
under this barrier. The quantum result depends strongly on the
mass defect of the point source.

The obtained in this article behavior of a quantum test particles
in the gravitational field of point source of gravity seems to us
to be much more physical then the one in the wide spread models of
black holes. Clearly, in contrast to such space-time holes with
nonphysical {\em infinitely} deep well in them, in our case the
finite inner well plays the role of a trap for the test particles.
It is not excluded that this way one may be able to construct a
model of very compact {\em matter} objects with an arbitrary large
mass and a size of order of Schwarzschild radius. It's possible
that such type of objects may explain the observed in the Nature
compact dark objects. They may be the final product of the stelar
evolution, instead of the quite formal and sophisticated
constructions like black holes which correspond to {\em empty
space} solutions of Einstein equations. Of course, at present
these possibilities are an open problem which deserves further
careful study.

\begin{figure}[ht]
    \begin{center}
        \includegraphics[totalheight=6cm,keepaspectratio]{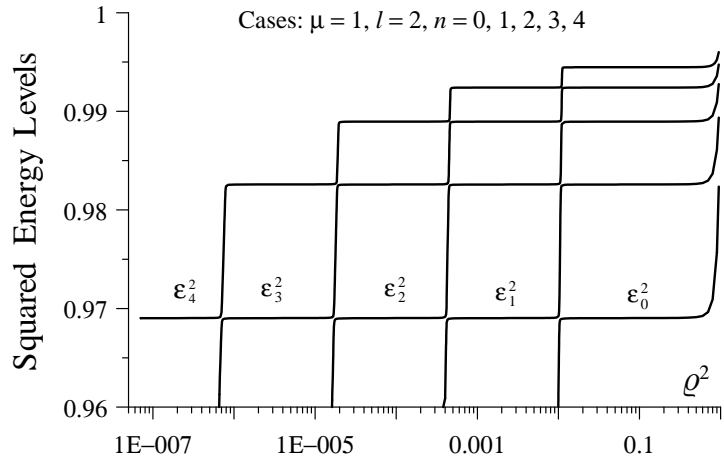}
        \caption{The attraction and the repulsion of the quantum energy levels
        for different values of the gravitational
        mass defect of the point source.}
        \label{F6}
    \end{center}
\end{figure}

It is easy to observe one more astonishing phenomena in the
discrete spectrum of a test quantum particle in gravitational
field of point source. It is related, too, with the mass defect of
the source. If one puts in the same figure the functional
dependence of all squared discrete energy levels
$\varepsilon^2_{n\,l}$ on the squared mass ratio
$\varrho^2=g_{tt}\!\mid_{r=0}$ (see in \cite{Fiziev}), one can
observe a repulsion and an attraction (up to a quasi-crossing) of
these levels, as shown in Fig.~\ref{F6}. Such type of behavior of
quantum discrete levels is well known in the laser physics, in the
neutrino oscillations and in other branches of quantum physics. To
the best of our knowledge we are observing such behavior of
quantum levels in the fundamental gravitational physics for the
first time.

\section{Conclusions}

In the present-days traditional approach to the motion of  test
particles in the Schwarzschild gravitational field, inside the
event horizon there exist a nonphysical infinitely deep potential
well. Everything which somehow can be placed inside this well will
fall unavoidable to its center $\rho=0$ for a finite time. It is
well known that this center is a physically inadmissible
geometrical singularity.  To hide such undesired singularity one
must use a unprovable mathematical hypothesis like the cosmic
censorship one. Today we have enough examples which show that such
hypotheses can not be correct at least in its original
formulation, see, for example \cite{nBH} and the references
therein.

In the article \cite{Fiziev} the principal role of the matter
point source of gravity was stressed. In GR the matter point
source presents a natural cutting factor for the {\em physical}
values of the luminosity variable $\rho\in [\rho_0, \infty)$,
where $\rho_0> \rho_G$ (\ref{rho_0}). This is because the infinite
mass density of the matter point changes drastically the geometry
of the space-time around it. In full accord with Dirac's
suggestion \cite{Dirac} this cutting places all nonphysical
phenomena, together with the very event horizon in the nonphysical
domain of the variables and yields a large number of new
interesting phenomena.

In particular, as shown in the present article, the described
cutting yields an interesting novel discrete spectra for
Klein-Gordon test particles in the gravitational field of massive
point. Mathematically this is caused by the universal change of
the boundary conditions for the corresponding wave equation, due
to the presence of the matter source of gravity and depends on its
mass defect.

A {\em real} discrete spectrum does not exist in the case of black
hole solution, just because of the presence of a hole in the
space-time, which absorbs everything around it. Our consideration
gives a unique possibility for {\em a direct experimental test} of
the existence of space-time holes. If one will find a discrete
spectra of corresponding stationary waves, propagating around some
compact dark object, one will have indisputable evidence that
there is no space-time hole inside such object. If, in contrast,
one will find only a decaying quasi-normal modes  with complex
eigenvalues (see \cite{QNM} and the references herein), this w ill
indicate that we are observing indeed a black hole.

We have obtained similar results for electromagnetic and
gravitational waves in the gravitational field of point
particle\cite{FBD}. For them the properly modified quasi-normal
modes turn to depend on gravitational mass defect, too. The
further study of these phenomena is an important physical problem
and will give us more practical physical criteria for an
experimental distinguishing of the two complete different
hypothetical objects: 1) the space-time black holes and 2) the
possible new very compact dark objects made of {\em real} matter.
Present days astrophysical observations are still not able do make
a difference between them. The only real observational fact is
that we are seen very compact and very massive objects, which show
up {\em only} due to their strong gravitational field. An actual
theoretical problem is to find a convincing model for description
of these already observed compact dark objects and the criteria
for the experimental justification of such model. We hope that the
present article is a new step in this direction.

\vskip .5truecm

{\em \bf Acknowledgments} \vskip .3truecm

One of the authors (PPF) is grateful to the High Energy Physics
Division, ICTP, Trieste, for the hospitality and for the nice
working conditions during his visit in the autumn of 2003. There
the idea of the present article was created. PPF is grateful to
the JINR, Dubna, too, for the priority financial support of the
present article. All of the authors are deeply indebted to the
JINR, Dubna for the hospitality and for the excellent working
conditions during their three months stay there at the end of 2003
and at the beginning of 2004, when the most of the work was done.
We wish to acknowledge, too, the participants in the scientific
seminars of the BLTF and LIT of JINR for the stimulating
discussions of the basic ideas, methods and results of the present
article.

\end{document}